\documentclass[lettersize,journal]{IEEEtran}
\usepackage{amsmath,amsfonts}
\usepackage{algorithmic}
\usepackage{array}
\usepackage[caption=false,font=normalsize,labelfont=sf,textfont=sf]{subfig}
\usepackage{textcomp}
\usepackage{stfloats}
\usepackage{url}
\usepackage{verbatim}
\usepackage{graphicx}
\def\BibTeX{{\rm B\kern-.05em{\sc i\kern-.025em b}\kern-.08em
    T\kern-.1667em\lower.7ex\hbox{E}\kern-.125emX}}
\usepackage{balance}
\usepackage{upgreek}
\usepackage{xurl}

\def\BibTeX{{\rm B\kern-.05em{\sc i\kern-.025em b}\kern-.08em
T\kern-.1667em\lower.7ex\hbox{E}\kern-.125emX}}
\begin{document}

\onecolumn
\thispagestyle{empty}
{\LARGE{\noindent \textbf{IEEE Copyright Notice} 
\\
\\
\\
\\
© 2023 IEEE. Personal use of this material is permitted. Permission from IEEE must be obtained for all
other uses, in any current or future media, including reprinting/republishing this material for advertising or promotional purposes, creating new collective works, for resale or redistribution to servers or lists, or reuse of any copyrighted component of this work in other works.
}}
\twocolumn
\setcounter{page}{1}

\title{Active Personal Eye Lens Dosimetry with the Hybrid Pixelated Dosepix Detector}
\author{Florian Beißer$^1$, Dennis Haag$^1$, Rafael Ballabriga$^2$, Rolf Behrens$^3$, Michael Campbell$^2$, Christian Fuhg$^3$, Patrick~Hufschmidt$^4$, Oliver Hupe$^3$, Carolin~Kupillas$^1$, Xavier Llopart$^2$, Jürgen~Roth$^3$, Sebastian Schmidt$^5$, Markus Schneider$^1$, Lukas Tlustos$^{2,6}$, Winnie Wong$^7$, Hayo~Zutz$^3$, and Thilo~Michel$^1$
\thanks{This work did not involve human subjects or animals in its research.}
\thanks{F. Beißer, D. Haag, C. Kupillas, M. Schneider, and T. Michel are with the Erlangen Centre for Astroparticle Physics, 91058 Erlangen, Germany (e-mail: florian.beisser@fau.de).}
\thanks{R. Ballabriga, M. Campbell, X. Llopart, and L. Tlustos are with CERN, 1211 Geneva, Switzerland}
\thanks{R. Behrens, C. Fuhg, O. Hupe, J. Roth, and H. Zutz are with Physikalisch-Technische Bundesantalt (PTB), 38116 Braunschweig, Germany}
\thanks{P. Hufschmidt was with the Erlangen Centre for Astroparticle Physics, 91058 Erlangen, Germany, and is now with Helene-Lange-Gymnasium, 90762 Fürth, Germany}
\thanks{S. Schmidt was with the Erlangen Centre for Astroparticle Physics, 91058 Erlangen, Germany, and is now with CodeCamp:N GmbH, 90429 Nuremberg, Germany}
\thanks{L. Tlustos is with the Institute of Experimental and Applied Physics, Czech Technical University in Prague, CZ}
\thanks{W. Wong was with CERN, 1211 Geneva, Switzerland. She is now with
Mercury Systems, 1212 Geneva, Switzerland.}
}

\maketitle


\begin{abstract}
Eye lens dosimetry has been an important field of research in the last decade. Dose measurements with a prototype of an active personal eye lens dosemeter based on the Dosepix detector are presented. The personal dose equivalent at 3\,mm depth of soft tissue, \(\boldsymbol{H}_\textbf{p}\boldsymbol{(3)}\), was measured in the center front of a water-filled cylinder phantom with a height and diameter of 20\,cm. The energy dependence of the normalized response is investigated for mean photon energies between 12.4\,keV and 248\,keV for continuous reference radiation fields (N-series) according to ISO 4037. The response normalized to N-60 ($\boldsymbol{\overline{E}=47.9\,\text{keV}}$) at \(\boldsymbol{0^\circ}\) angle of irradiation stays within the approval limits of IEC 61526 for angles of incidence between \(\boldsymbol{-75^\circ}\) and \(\boldsymbol{+75^\circ}\). Performance in pulsed photon fields was tested for varying dose rates from \(\boldsymbol{0.1\,\frac{\textbf{Sv}}{\textbf{h}}}\) up to \(\boldsymbol{1000}\,\frac{\textbf{Sv}}{\textbf{h}}\) and pulse durations from \(\boldsymbol{1\,\textbf{ms}}\) up to \(\boldsymbol{10\,\textbf{s}}\). The dose measurement works well within the approval limits (acc. to IEC 61526) up to \(\boldsymbol{1\,\frac{\textbf{Sv}}{\textbf{h}}}\). No significant influence of the pulse duration on the measured dose is found. Reproducibility measurements yield a coefficient of variation which does not exceed \(\boldsymbol{1\,\%}\) for two tested eye lens dosemeter prototypes.
\end{abstract}


\section{Introduction}
\label{sec:introduction}
\IEEEPARstart{T}{he} lens of the eye is susceptible to radiation induced damages such as a significant increase of the likelihood for cataractogenesis \cite{icrp118}\cite{Chodick}\cite{Nakashima}\cite{Neriishi}.
Eye lenses of average persons who are not occupationally exposed to ionizing radiation, apart from cosmic background radiation and other natural exposure, will not experience significant harm apart from non-radiation causes such as age, eye injuries, diabetes etc.
However occupationally exposed personnel e.g. in interventional radiology and cardiology can be particularly affected.
Vano et al. \cite{Vano} investigated the influence of radiation safety equipment such as lead glass goggles and shielding on the received eye lens dose.
Without protective equipment, up to 3.72\,mSv per procedure could be achieved in worst-case scenarios.
Radiation protection goggles or a lead shield reduced the received dose down to about 0.1\,mSv.
In a complementary study \cite{Vano2}, posterior subcapsular lens changes in operating medical staff were examined.
Effects were found for 50\,\% of interventional cardiologists and 41\,\% of nurses with accumulated doses of 0.1\,Sv to 18.9\,Sv.
The recommended dose limit for the lens of the eye has been tightened in 2011 by the \textit{International Commission on Radiological Protection} (ICRP)~\cite{icrp118}.
An average dose of \(20\,\frac{\text{mSv}}{\text{a}}\) must not be exceeded within a 5-year-period.
Additionally, the dose within a single year must never exceed 50\,mSv.
These limits were adopted by the International Atomic Energy Agency (IAEA)~\cite{iaea}, European Union~\cite{euratom_rolf} and many national legislations, e.g., in Germany~\cite{Strahlenschutzgesetz}.
Therefore, monitoring systems for the dose to the lens of the eye are necessary in order to assure that the dose limit is not exceeded.

The personal dose equivalent \(H_\text{p}(3)\) at a depth of 3\,mm in soft tissue, defined by the \textit{International Commission on Radiation Units and Measurements} (ICRU)~\cite{icru}\cite{icru2}, is utilized to estimate the dose to the lens of the eye \cite{Dos23}.

To date (April 2023), there is no legally approved active eye lens dose monitoring system and only two passive dosemeters available \cite{mirion}\cite{lps}.
The major advantage of active dosemeters is the real-time availability of the dose information allowing for direct countermeasures if necessary.

In this work, first measurements with a prototype of an active eye lens dosemeter based on the hybrid pixel detector Dosepix are presented.

\section{The Dosepix Detector}

Dosepix \cite{Winnie_Paper} is a hybrid energy-resolving dead-time free photon-counting pixelated x-ray detector.
It was developed by a collaboration of Friedrich-Alexander-Universität Erlangen-Nürnberg (FAU) and the European Organization for Nuclear Research CERN.
\begin{figure}
    \centerline{\includegraphics[width=3.5in]{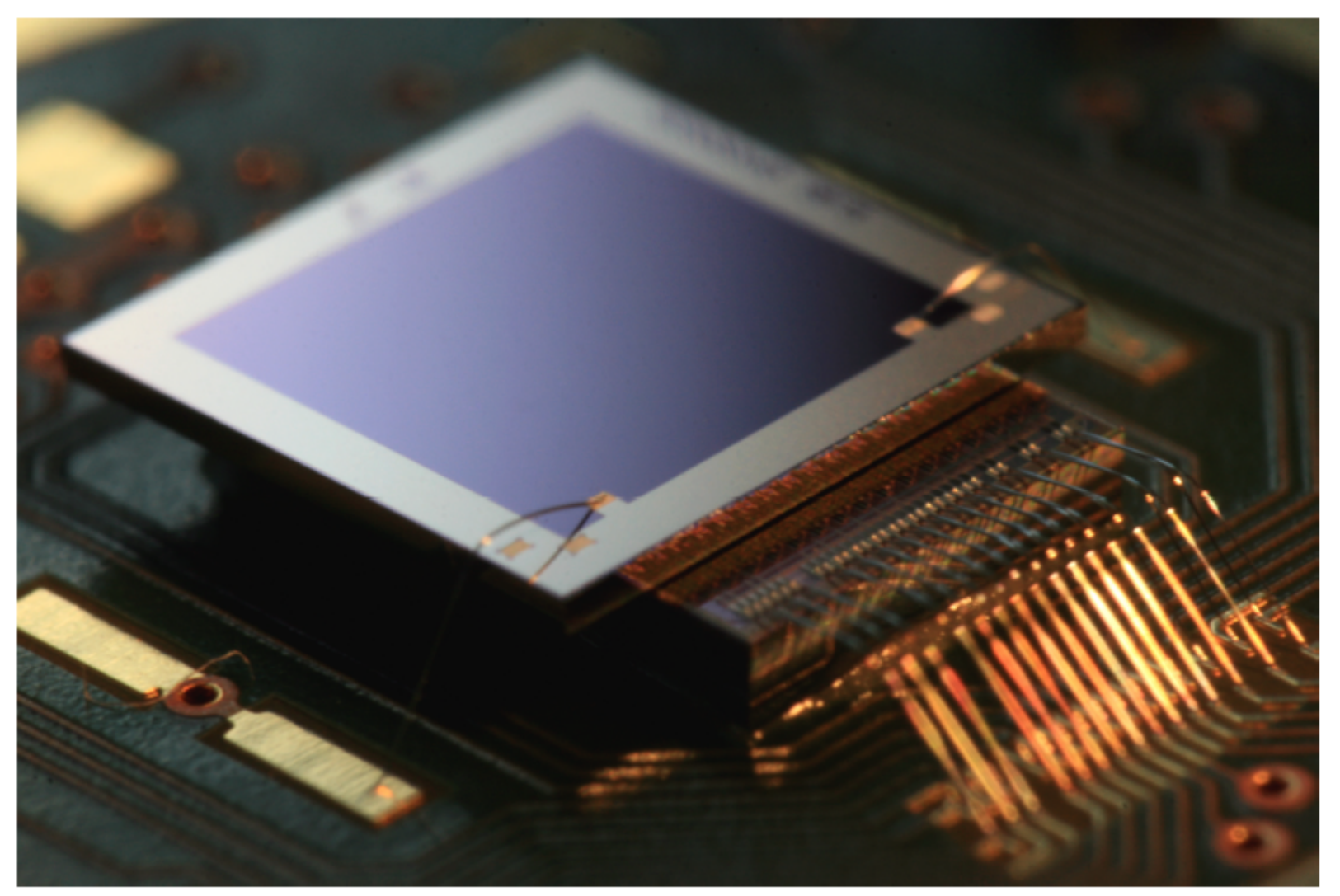}}
    \caption{Photograph of the Dosepix detector with a $300\,\upmu$m thick p-in-n doped silicon sensor layer attached. The sensor is surrounded by a guard ring. Wire bonds in the bottom right of the photograph connect Dosepix to read-out hardware. The picture is taken from \cite{Gabor}.}
    \label{fig:Dosepix}
\end{figure}
Detailed information about the detector can be found in \cite{Winnie_Paper}\cite{Winnie}\cite{Zang}\cite{Ritter_Paper}.
Dosepix is of hybrid design meaning that its \textit{application specific integrated circuit} (ASIC) allows bump-bonding to different sensor layers depending on the use case.
For the purpose of eye lens dosimetry, a \(300\,\upmu\)m thick p-in-n doped silicon sensor is used.
A photograph of Dosepix bump-bonded onto such a sensor layer is shown in \figurename\ref{fig:Dosepix}.
A bias voltage of 48\,V is applied to drift charge carriers to the pixel electrodes.
This bias voltage is sufficient to fully deplete the sensor \cite{Maggus}.
It is doped such that the whole sensor is divided into \(16\times16\) pixels of square shape with a pixel pitch of \(220\,\upmu\)m.
The pixels in the upper two and lower two rows have an edge length of \(55\,\upmu\)m for processing higher dose rates.
Therefore, there are non-sensitive areas between the small pixels where charge carriers are not transported towards a pixel electrode.
The pixels in the middle 12 rows with an edge length of \(220\,\upmu\)m are flush.
All pixels combined result in an overall sensitive area of \(9.49\,\text{mm}^2\).
This work focuses exclusively on results for the large pixels as they are sufficient by themselves to fulfill all of the tested approval requirements.
The detector has a lower energy threshold of about 10\,keV.
Every deposited energy signal is processed by the electronics of the respective pixel and sorted into a 16-channel energy histogram with freely adjustable bin edges.
These histograms are used to calculate the eye lens dose.
The lowest energy bin edge corresponds to 12\,keV.
This means that deposited energies between \(\approx10\,\)keV and \(12\,\)keV are not stored in the final histograms because they did not proof to be well suited for dosimetry as they show a poor energy resolution and are disturbed by noise events.
The last bin serves as an overflow bin and contains all registered events with energy depositions above 150\,keV.
Dosepix has no read-out dead-time which is realized by reading out the energy histograms of each pixel column individually one after another while the remaining pixels continue to process signals.

The total numbers of entries \(N_i\) in the i-th energy bin, summed over the event histograms of all pixels, are used to calculate the dose.
This is achieved via a weighted sum of all histogram entries where the weights are given by dose conversion factors \(k_i\).
Thus, the dose equivalent measured by Dosepix \(H_\text{p}(3)_\text{DPX}\) is given by:
\begin{equation}
    H_\text{p}(3)_\text{DPX}=\sum_{i=1}^{16}k_i N_i.
    \label{eq:KonvFak}
\end{equation}
The conversion factors \(k_i\) are determined by a least-squares fitting algorithm via a combination of simulation data of monoenergetic photon beams and measurements at reference conditions according to the procedure described in \cite{Bastl}.
By this design, the conversion factors \(k_i\) represent the physical processes of the system.
This method has previously been utilized successfully to create an \(H_\text{p}(10)\) dosemeter based on three Dosepix detectors which works well in continuous and pulsed radiation fields~\cite{Deniz1}\cite{Deniz2}.

\section{Methods}

All measurements were performed at the facilities of Physikalisch-Technische Bundesanstalt (PTB).
The relevant dose quantity for estimating the eye lens dose is the personal dose equivalent \(H_\text{p}(3)\).
A phantom is required to replace the human head in order to establish reference conditions for \(H_\text{p}(3)\) irradiations.
The corresponding phantom was suggested by working group~2 of the \textit{Optimization of Radiation Protection of Medical Staff} (ORAMED) project and is designed to approximate scattering behavior of a human head~\cite{Bordy}\cite{iso4037_3}.
It is a water-filled \textit{polymethyl methacrylate} (PMMA) cylinder with equal height and diameter of 20\,cm.
The walls have a thickness of 5\,mm. 
The eye lens dosemeter prototype was placed on the center front of the phantom.
The response $R$ is defined as the ratio between the dose determined by Dosepix and the reference dose, the latter being measured using a previously calibrated monitor chamber installed at the utilized x-ray facilities.
This monitor calibration is traceable to the international systems of units (SI) \cite{si} using PTB’s corresponding primary standards.
Its systematic dose uncertainty is as small as 2\,\% (one sigma), therefore considered to be negligible for this work. 
Approval limits for dose influencing quantities are stated for the normalized response which is given by the response, normalized to the response of a chosen set of influencing parameters.
In this work, the normalization was performed with respect to the response at continuous irradiation of N-60 at \(0^\circ\) angle of radiation incidence.

The prototype of an active eye lens dosemeter consists of a Dosepix detector positioned in front of read-out hardware inside a 3D-printed case.
\begin{figure}
    \centerline{\includegraphics[width=3.5in]{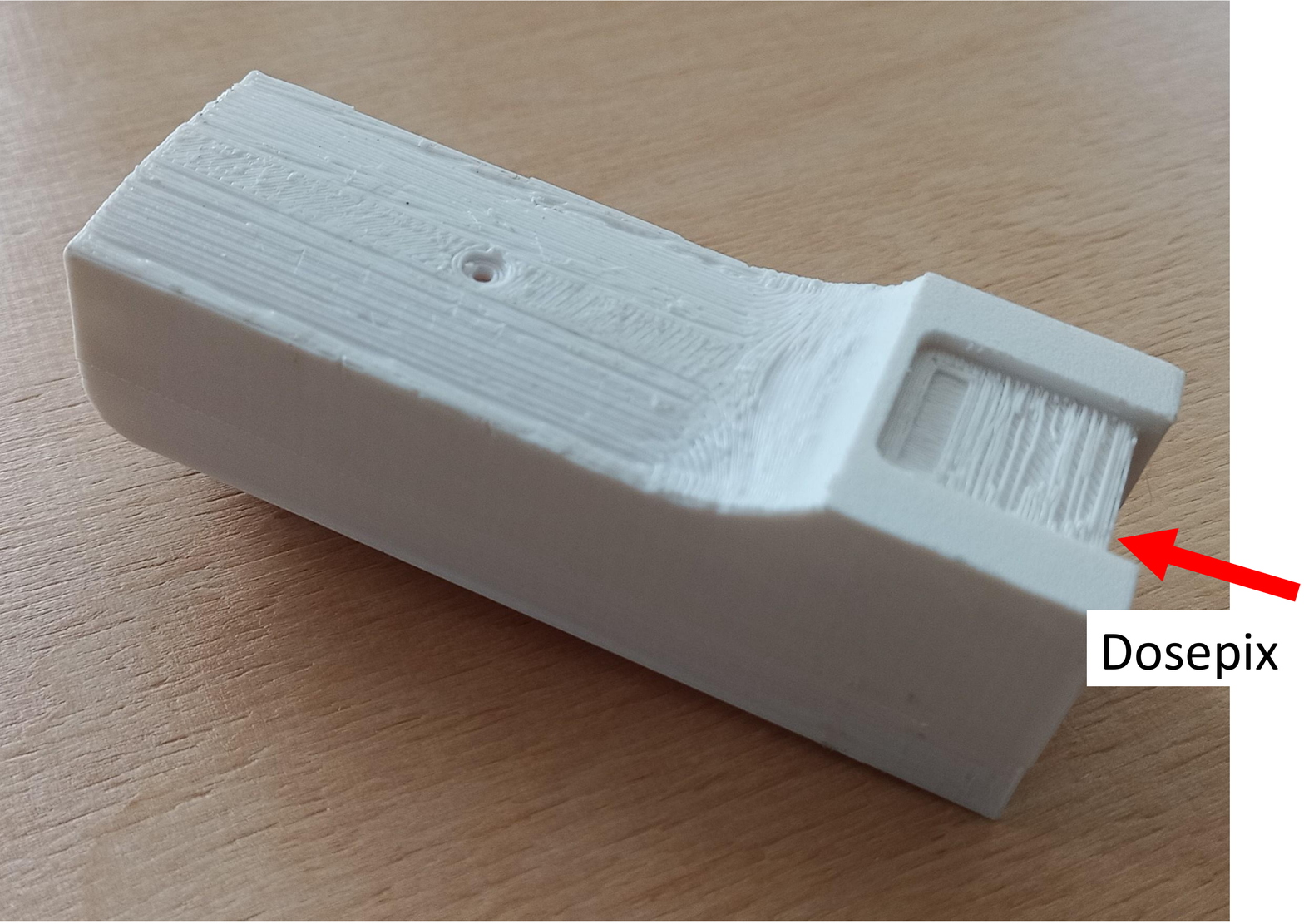}}
    \caption{Photograph of the active eye lens dosemeter prototype. The red arrow indicates the position and alignment of the installed Dosepix detector.}
    \label{fig:ELD}
\end{figure}
\begin{figure}
    \centerline{\includegraphics[width=3.5in]{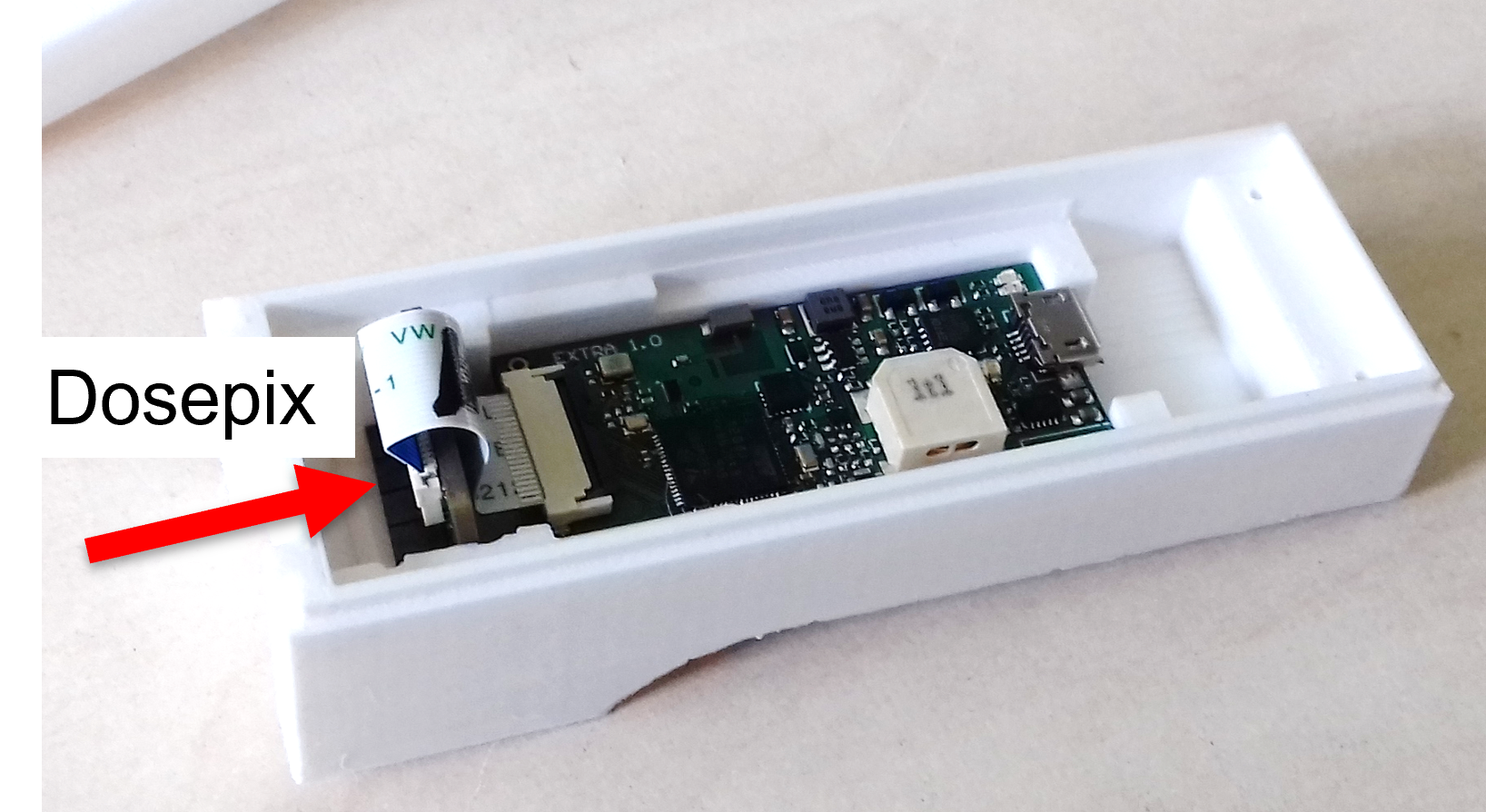}}
    \caption{Photograph of the read-out hardware inside the dosemeter case. The red arrow indicates the position and alignment of the installed Dosepix detector.}
    \label{fig:ELD_offem}
\end{figure}
Photographs of the dosemeter are shown in \figurename\ref{fig:ELD} (closed case) and \figurename\ref{fig:ELD_offem} (open with read-out hardware).
Its wall has a thickness of about 2\,mm and is made of \textit{Acrylnitril-Butadien-Styrol-Copolymer} (ABS).
Dosepix is positioned about 2.5\,mm behind the case wall with a PMMA ring surrounding the detector ensuring mechanical stability and providing stability against sudden acceleration of the device which might push Dosepix against the wall destroying the sensor layer.
This position of Dosepix inside the case is used as the reference point with regard to rotations and distances. 
The detector is connected via flat cable to the main hardware.
The latter is built around a microcontroller which manages voltage and current supplies and is responsible for all communication between a PC and Dosepix.
The final eye lens dosemeter will be powered by a battery and data being transferred via Bluetooth.
However, the current prototype is charged and read-out via a USB-connection at the back of the dosemeter.
The case itself has a length of 8.9\,cm and a front face of $3.0\,\text{cm}\times2.8\,\text{cm}$.
This is small enough to allow the dosemeter to be worn on the side of the head using a headband and is acceptable for investigation measurements identifying possible shielding or procedure improvements.
With these improved protection measures, routine use of such a large dosemeter is not necessary.


The voltage range of the utilized x-ray tubes goes from 15\,kV to 300\,kV.
Dosemeter rotation is executed with respect to the reference point mentioned earlier.
The radiation qualities of the N-series according to ISO 4037-1 \cite{iso4037_1} were used for the majority of the performed irradiations.
These photon radiation qualities are designed to be energetically narrow compared to other radiation series from \cite{iso4037_1}.
Additional radiation qualities from W- and H-series (acc. to \cite{iso4037_1}) were added in the low energy regime \(\leq 40\,\text{keV}\) as dependencies on the mean photon energy and the energy width are expected especially in this energy regime.
For pulsed measurements (irradiation pulse durations \(\leq 10\,\)s according to~\cite{iso18090_1}), the radiation qualities RQR-5 and RQR-8 \cite{iec61267} were both used to access information on the energy dependence of the dose rate dependence. 
The majority of the measurements were performed at a distance of 2.5\,m between the x-ray tube and the reference point of the dosemeter to ensure a homogeneous irradiation of the dosemeter and the water cylinder phantom.
If more measurement repetitions of a radiation quality with the same parameters were performed, the statistical uncertainty of the dose calculation was derived from the standard deviation of the individual dose values. 
Otherwise, statistical uncertainty is estimated from Poisson statistics where the standard deviation \(\sigma_\text{N}\) of the number of registered counts \(N\) is given by \(\sigma_\text{N}=\sqrt{N}\) according to \cite{gum}.
Thus, the standard deviation of the dose \(\sigma_{H_\text{p}(3)}\) is approximately given by
\begin{equation}
    \sigma_{H_\text{p}(3)}=\sqrt{\sum_{i=1}^{16}k_i^2N_i}
    \label{eq:Poisson}
\end{equation}
resulting from the law of uncertainty propagation assuming independent \(N_i\) which only holds true as an approximation.


\section{Simulation and Dose Conversion Factor Determination}

A Monte-Carlo simulation of the eye lens dosemeter was performed prior to irradiations at facilities of PTB.
Detailed descriptions about the utilized simulation chain are given in \cite{Bastl}.
Simulation results are necessary for the determination of the dose conversion factors, introduced in Eq. \ref{eq:KonvFak}.
The simulation setup included a model of the active eye lens dosemeter prototype, consisting of Dosepix, read-out hardware and the dosemeter case.
These parts were positioned in the center front of the previously described water cylinder phantom.
The simulation itself consisted of two parts to reduce the total computation time.
At first, the backscattering behavior of the phantom under homogeneous irradiation was simulated.
For this, the simulation toolkit Geant~4 version~10.7.2 was utilized~\cite{geant}.
The setup for this first simulation step consisted of the phantom surrounded by a ring-shaped ideal photon detector with perfect sensitivity existing only for simulation purposes.
This detector saves energy, detection position and incidence direction of all photons that previously interacted with the phantom.
It was positioned at a height of 9.25\,cm with respect to the bottom of the phantom.
The detector has an inner diameter of 20\,cm, an outer diameter of 40\,cm and a height of 4\,cm.
Thus, this detector covers the entire volume that is claimed by the dosemeter.
Backscattered photons are detected with perfect energy and spatial resolution whereas the detector is transparent with regards to photons from the primary beam.
A square source shape of 20\,cm x 20\,cm with a parallel beam irradiating the complete front face of the cylinder water phantom was chosen.
Energies from 10\,keV up to 250\,keV in steps of 1\,keV were each simulated with a total number of $5\cdot10^8$ photons.
From this, the influence of the phantom on the photon spectrum at the position of Dosepix was estimated.
Normally, a simulation of an \(H_\text{p}(3)\)-setup would require a homogeneous irradiation of the entire cylinder phantom.
However, the performed phantom backscattering simulation allows for the main simulation to only consider the front area of the dosemeter while receiving all remaining necessary information from the former.
This drastically reduces the total computation time.

The main simulation of the sensor of Dosepix was based on the Monte Carlo simulation toolkit Allpix$^2$ \cite{Allpix}.
As written above, monoenergetic photon beams with energies between 10\,keV and 250\,keV in steps of 1\,keV were simulated to match the backscattering simulation.
Only the charge propagation and charge sharing between different pixels in the sensor layer were simulated.
The pixel electronics here were assumed to be ideal and noise free.
The Dosepix electronics and the read-out hardware were only considered in terms of scattering effects.
\begin{figure}
    \centerline{\includegraphics[width=3.5in]{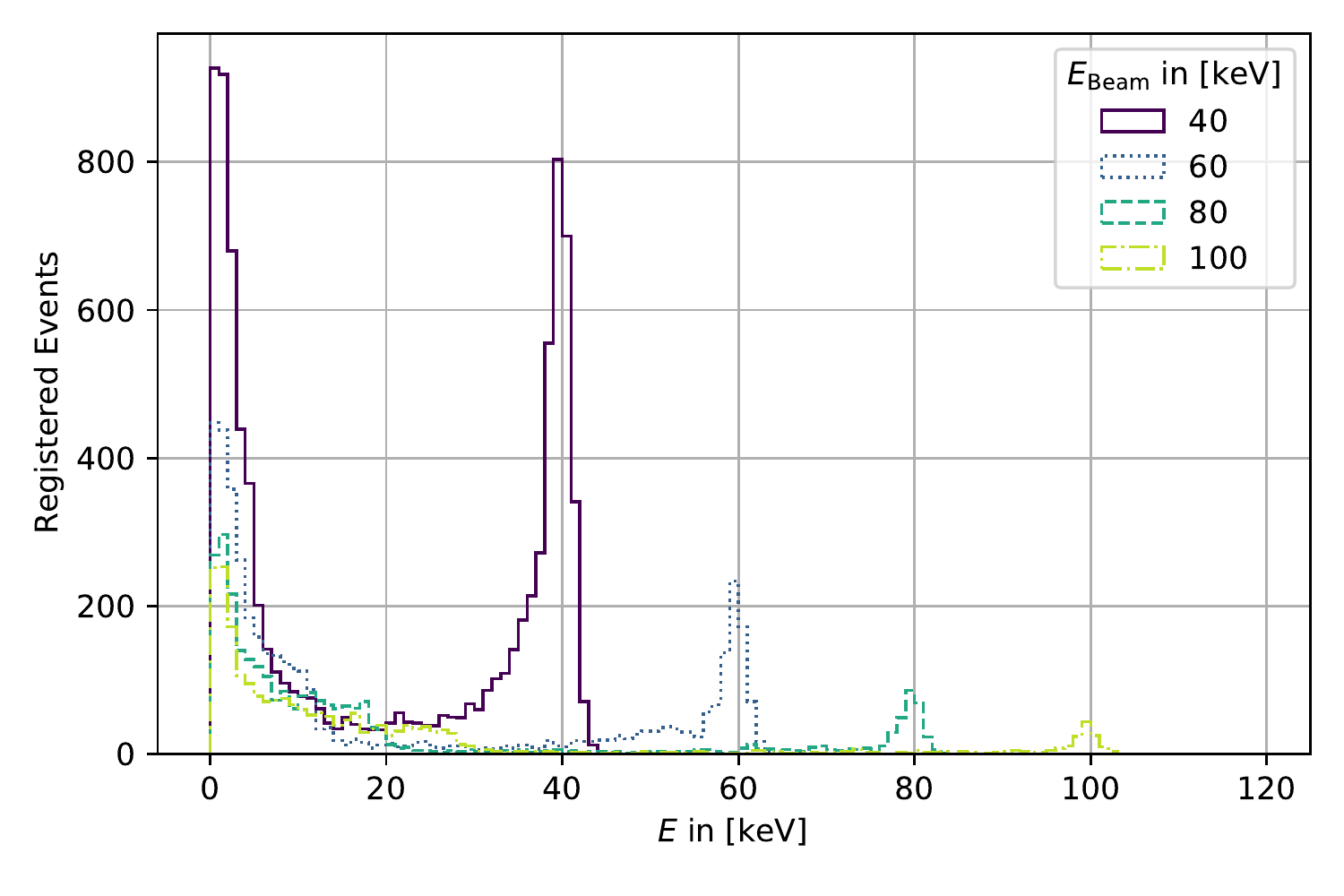}}
    \caption{Simulated sensor signal of the large pixels of Dosepix as a part of the eye lens dosemeter prototype in front of the cylinder phantom. $5\cdot10^8$ monoenergetic photons with a primary beam energy \(E_\text{Beam}\) were irradiated homogeneously onto the phantom area.}
    \label{fig:SimSpek}
\end{figure}
Some exemplary simulated energy deposition spectra are shown in \figurename \ref{fig:SimSpek}.
Full-energy peaks and Compton continua are the dominating features in the spectra.
The interaction cross section of photons and the sensor material via Photoelectric effect decreases with increasing photon energy which is reflected by the decreasing number of events in the full-energy peaks \cite{Bethge}.

These simulated spectra were re-binned into the same 16 energy bins used for the performed measurements.
This mimics the data processing of the pixel electronics of an ideal noise free multichannel analyzer.
Therefore, dose values were calculated from simulation data as they would be in reality and for use as reference value for spectra measured by the real Dosepix detector. 
The respective reference dose is calculated from the photon fluence via conversion coefficients derived from \cite{icru57} and \cite{Gual_Ka_Hp3}.
Thus, the dose conversion factors can be optimized with respect to the reference dose.


\section{Results}

\subsection{Continuous Photon Fields}
\begin{figure}
    \centerline{\includegraphics[width=3.5in]{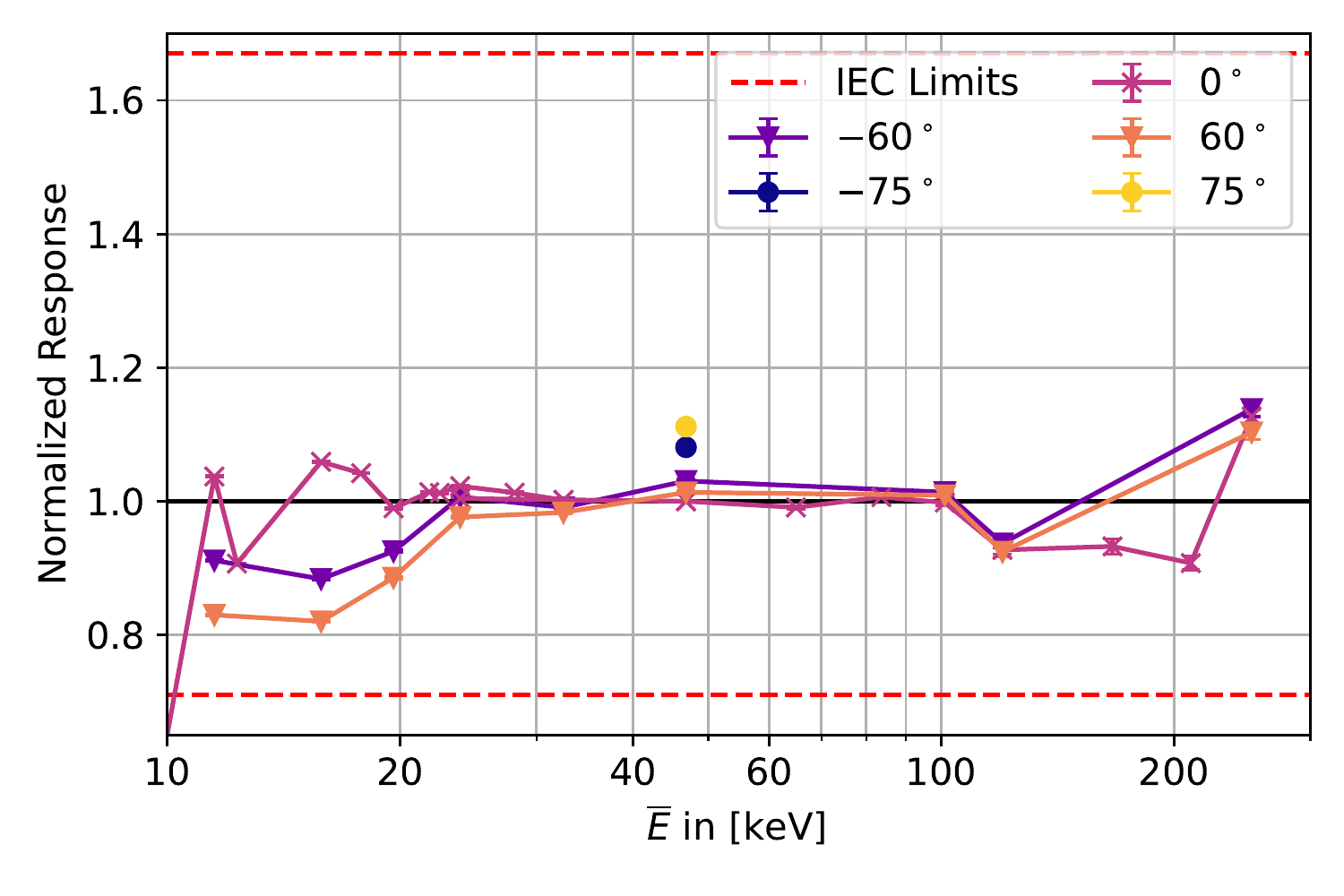}}
    \caption{Normalized response of the active eye lens dosemeter prototype (number 1) with respect to the mean energy of the irradiated photons for different angles of radiation incidence. The data points at \(0^\circ\) were used for determining the dose conversion factors in Eq.\ref{eq:KonvFak}. Error bars represent statistical uncertainties only and are hardly visible. The red dashed lines indicate the limits stated in IEC 61526~\cite{iec61526}, the black line represents an ideal response of 1.0.}
    \label{fig:EAbh}
\end{figure}
The combined influence of mean photon energy and angle of radiation incidence was investigated in the following.
For this purpose, reference radiation fields of the N-series with a reference dose of $167\,\upmu$Sv were utilized.
This value approximately lies in the middle of the low dose regime in terms of dose variation ($H_\text{p}(3)<300\,\upmu$Sv acc. to \cite{iec61526}).
Also, $167\,\upmu$Sv was chosen internally in the past for dosimetry measurements with Dosepix and were reused for comparability reasons. 
Additionally, radiation qualities W-30, W-40, H-20, H-30, and H-40 were used for 0° angle of radiation incidence.
\figurename~\ref{fig:EAbh} shows the energy and angular dependence of the response normalized to (N-60, 0°). 
The normalized response of the dosemeter prototype is depicted with respect to the mean irradiated energy \(\overline{E}\) for different angles of radiation incidence.
Statistical uncertainties are estimated from Poisson statistics via Eq. \ref{eq:Poisson}.

The N-series measurements at $0^\circ$ were used for the determination of the dose conversion factors together with the simulation data mentioned above.
Therefore, the results in \figurename~\ref{fig:EAbh} indicate the optimum response of the presented eye lens dosemeter prototype.
Dashed red lines indicate the approval limits for the normalized response of 0.71 to 1.67 stated in \cite{iec61526} and \cite{ptba23e}.
The normalized response stays within the approval limits throughout the entire tested mean energy range \(\overline{E}=12.4\,\text{keV}\) to \(\overline{E}=248\,\text{keV}\) (N-15 to N-300).
Dose estimation start to worsen for radiation qualities with higher mean energies \(>100\,\)keV because, as already mentioned, events with deposited energies above 150\,keV are all stored in the same bin.

For lower energies, the response for \(\pm 60^\circ\) is  lower than for \(0^\circ\) whereas there is no such significant difference for higher photon energies.
The reason is that larger absolute angles of incidence result in an increased effective thickness of the dosemeter housing.
Therefore, more photons are absorbed before reaching the detector and less events are deposited in the sensor of Dosepix.
This effect decreases for higher photon energies because the interaction cross section is smaller for such photons \cite{Bethge}.
An exemplary measurement of the radiation quality N-60 at \(\pm75^\circ\) still results in a response well within the IEC 61526 limits \cite{iec61526}.
At this angle of radiation incidence, photons hit the edge of the front face of the dosemeter case and its side wall.

A second run of measurements was performed half a year later with two prototypes.
Prototype 1 is the same device as presented as before.
The discharge current \(I_\text{Krum}\) (see. \cite{Winnie}) was changed from 2.2\,nA to 6.6\,nA which required a new energy calibration.
This was done because a higher \(I_\text{Krum}\) was shown to reduce the temperature dependence of Dosepix \cite{self.bachelor}.
Prototype 2 is an almost identical device, the main difference being the electronics adapter, connecting Dosepix to the read-out hardware, is positioned right behind the Dosepix detector in prototype 1, whereas it is shifted about 3\,mm in a lateral direction for prototype 2.
This requires the introduction of a small hole measuring 0.7\,cm by 1.5\,cm into the dosemeter case of prototype 2 behind the detector which is covered with PMMA.
The reason for this difference was the unavailability of the exact hardware as it was used for prototype 1 at the time of assembly of prototype 2.
For the complementary measurements, the dose conversion factors are not changed for any of the detectors compared to the first measurements presented above.
N-series radiation qualities from N-20 to N-300 were measured for the selected angles of radiation incidence between \(-75^\circ\) and \(+75^\circ\).
Not every radiation quality was measured for each angle due to a limited availability of the x-ray facilities.
\begin{figure}
    \centerline{\includegraphics[width=3.5in]{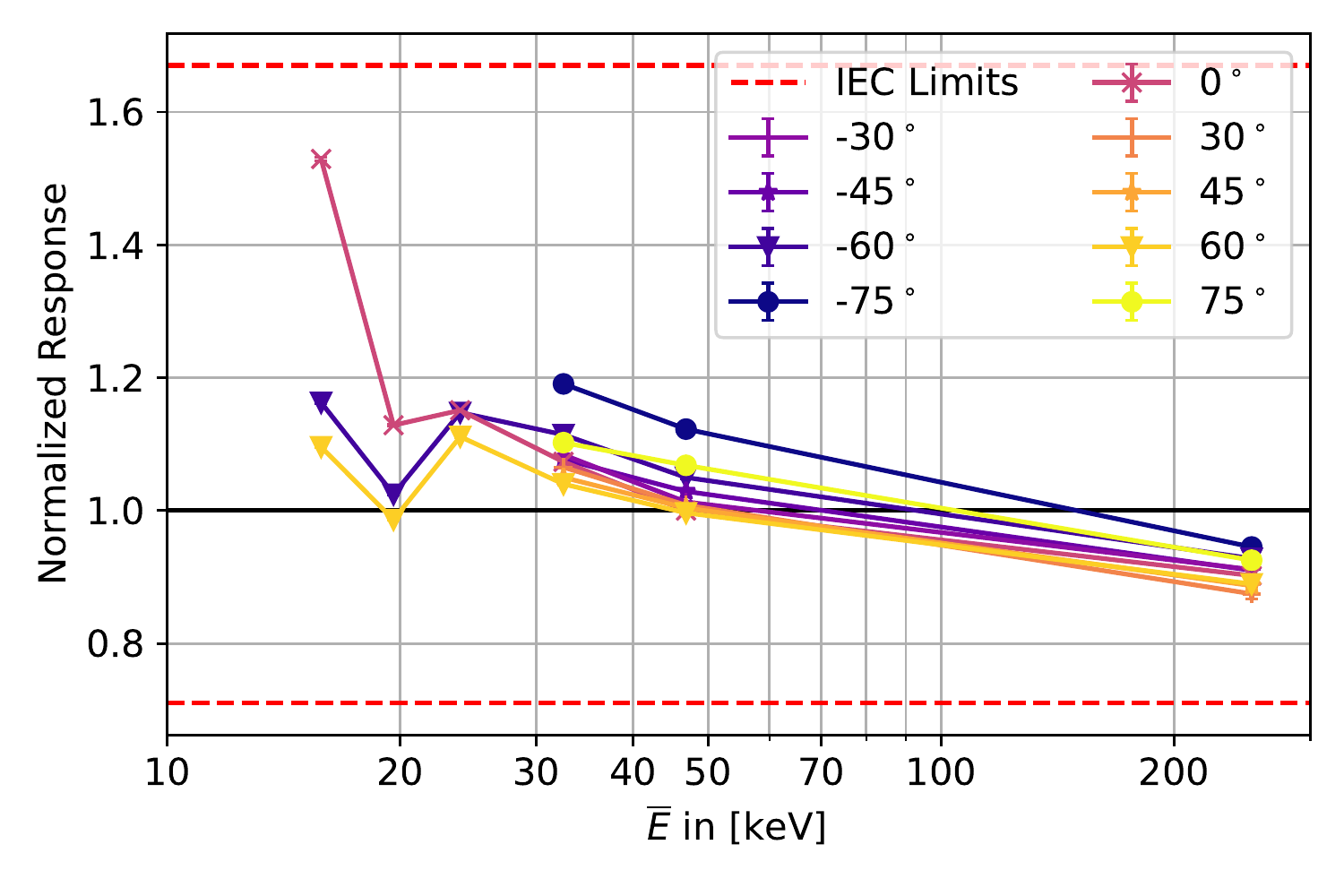}}
    \caption{Normalized response of the first active eye lens dosemeter prototype 1 (six months after the measurement for \figurename~\ref{fig:EAbh}) for selected radiation qualities and angles of radiation incidence. Error bars represent the estimated statistical uncertainties and are hardly visible in the plot. The red dashed lines indicate the limits stated in IEC 61526 \cite{iec61526}, the black line represents an ideal response of 1.0.}
    \label{fig:EnWi_relo1}
\end{figure}
\begin{figure}
    \centerline{\includegraphics[width=3.5in]{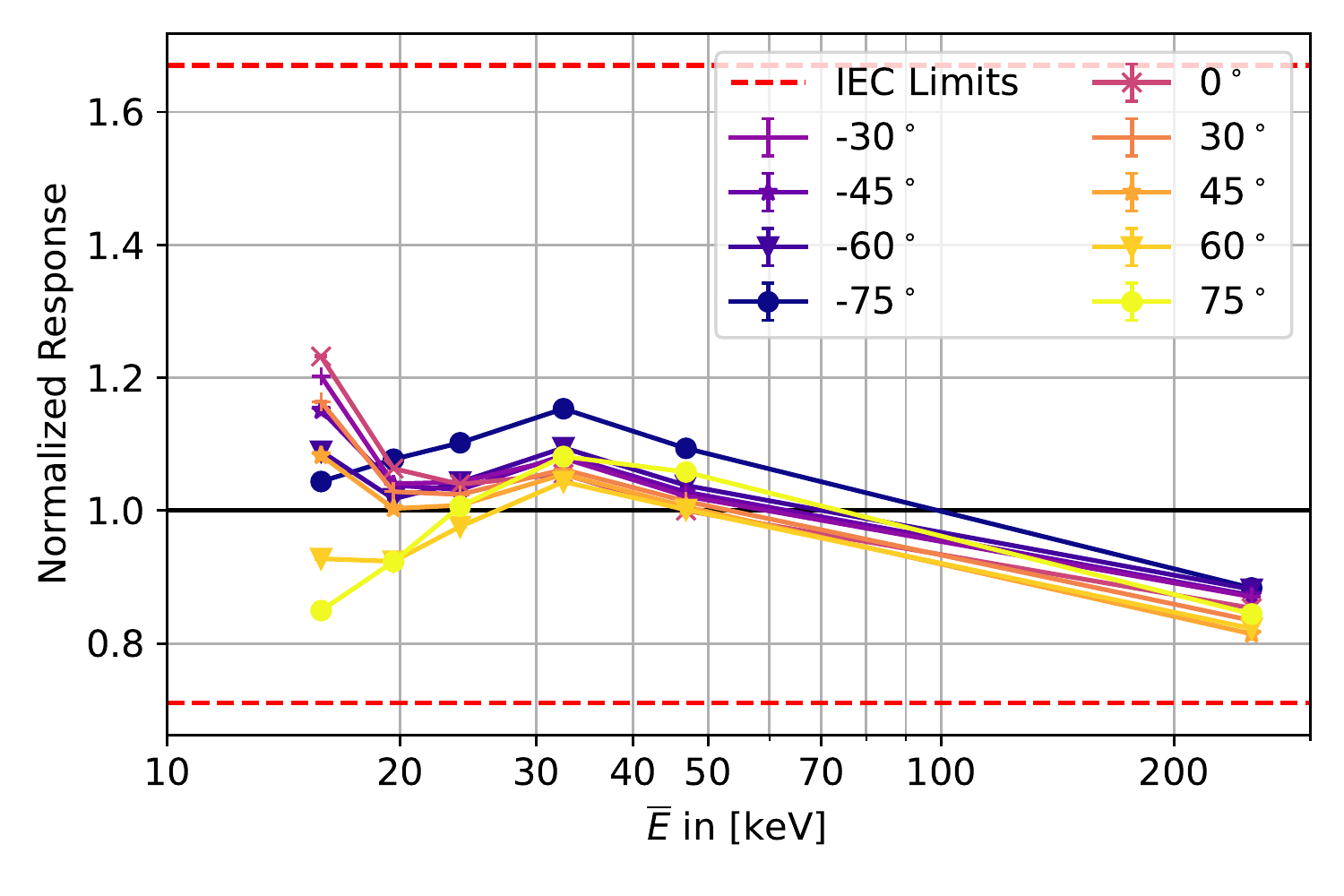}}
    \caption{Normalized response of the second active eye lens dosemeter prototype~2 (six months after the measurement for \figurename~\ref{fig:EAbh}) for selected radiation qualities and angles of radiation incidence. Error bars represent the estimated statistical uncertainties and are hardly visible in the plot. The red dashed lines indicate the limits stated in IEC 61526 \cite{iec61526}, the black line represents an ideal response of 1.0.}
    \label{fig:EnWi_relo2}
\end{figure}
The results are shown in \figurename~\ref{fig:EnWi_relo1} (prototype~1) and \figurename~\ref{fig:EnWi_relo2} (prototype~2).
Normalization was performed individually for each prototype with respect to the response at the radiation quality N-60 (\(\overline{E}=47.9\,\)keV) at $0^\circ$ angle of radiation incidence.
The absolute response values are $0.944\pm0.014$ for prototype 1 and $1.101\pm0.015$ for prototype 2
where the given uncertainties represent the statistical uncertainties according to Eq.\ref{eq:Poisson} neglecting the uncertainty of the reference dose value.
Red dashed lines indicate the limits of the normalized response 0.71 and 1.67 as stated in \cite{iec61526}\cite{ptba23e}.
All data points stay well within the limits.
The energy dependent normalized response is similar for both tested prototypes although there is a significant overestimation of the dose by prototype 2 compared to prototype 1 indicated by the given normalization factors.
Nonetheless, the overall trend indicates the possibility of a correction based on a simple proportionality factor.
Therefore, it might be possible to intercalibrate different eye lens dosemeters via measurements of few selected radiation qualities in the future.
The difficulty of estimating the dose for N-20 probably results from the spectrum being close to the low energy end of the sensitivity range (12\,keV) and the entire spectrum falling into two energy bins.
The higher normalized response of N-20 compared to the first set of measurements (\figurename\ref{fig:EAbh}) primarily results from the different energy calibration due to the higher \(I_\text{Krum}\) as the energy calibration is most sensitive to small energy shifts in the vicinity of the energy threshold \cite{Jakubek_cali}.
A higher angle of radiation incidence results in a smaller response because of the reasons mentioned before.
This second run of measurements shows that the active eye lens dosemeter prototype is capable of adequately measuring the dose up to $\pm 75^\circ$ surpassing the approval limits of $\pm 60^\circ$ stated in \cite{iec61526} and \cite{ptba23e}.
However, the comparison of the eye lens dosemeter prototypes is not completely exhaustive as no measurements of prototype 1 exist for mean energies \(\leq 30\,\)keV and angles of incidence of \(\pm 30^\circ\),\(\pm 45^\circ\), and \(\pm 75^\circ\).
Although no big changes of the normalized response are expected in this parameter space as indicated by the data points corresponding to angles of incidence of \(\pm 60^\circ\), this should be addressed in future investigations.

\subsection{Pulsed Photon Fields}
Dose measurements in pulsed photon fields were performed by irradiating the dosemeter prototype 1 with the radiation qualities RQR-5 (\(\overline{E}=40.3\,\)keV) and RQR-8 (\(\overline{E}=50.8\,\)keV) with varying dose rates \(\Dot{H}_\text{p}(3)\).
Each measurement is repeated four times.
The distance was 2.5\,m for \(\Dot{H}_\text{p}(3)\leq 1\,\frac{\text{Sv}}{\text{h}}\) (200\,ms pulses) and \(\Dot{H}_\text{p}(3)\leq 4\,\frac{\text{Sv}}{\text{h}}\) (1\,ms pulses).
For larger dose rates, the distance was reduced to 0.5\,m due to technical reasons.
A corresponding correction is applied to all measurement results performed at the smaller distance.
This correction accounts for the five times smaller beam diameter (due to the five times reduced distance) and omission of the water cylinder phantom as it would not have been fully irradiated at the smaller distance.
The latter was deduced from a measurement with and without the phantom.
\begin{table}
    \centering
    \caption{Dose correction factors multiplied at the data, presented in \figurename~\ref{fig:Rate}, for dose rates $>4\,\frac{\mathrm{Sv}}{\mathrm{h}}$ (\(1\,\mathrm{ms}\) pulse duration) or $>1\,\frac{\mathrm{Sv}}{\mathrm{h}}$ (\(200\,\mathrm{ms}\) pulse duration).}
    \begin{tabular}{|c|c|c|}
        \hline
        Radiation Quality &  1\,ms & 200\,ms\\
        \hline
         RQR-5 & $0.881\pm0.026$ & $0.973\pm0.011$ \\
         RQR-8 & $0.93\pm0.04$ & $0.95\pm0.05$ \\
         \hline
    \end{tabular}
    \label{tab:KorrFak}
\end{table}
The correction factors are listed in Table~\ref{tab:KorrFak}.
Their uncertainties are not considered in terms of the statistical uncertainties, drawn in the plots. 
One measurement series is performed with pulses of $200\,$ms duration and one with $1\,$ms pulses.
\begin{figure}
    \centerline{\includegraphics[width=3.5in]{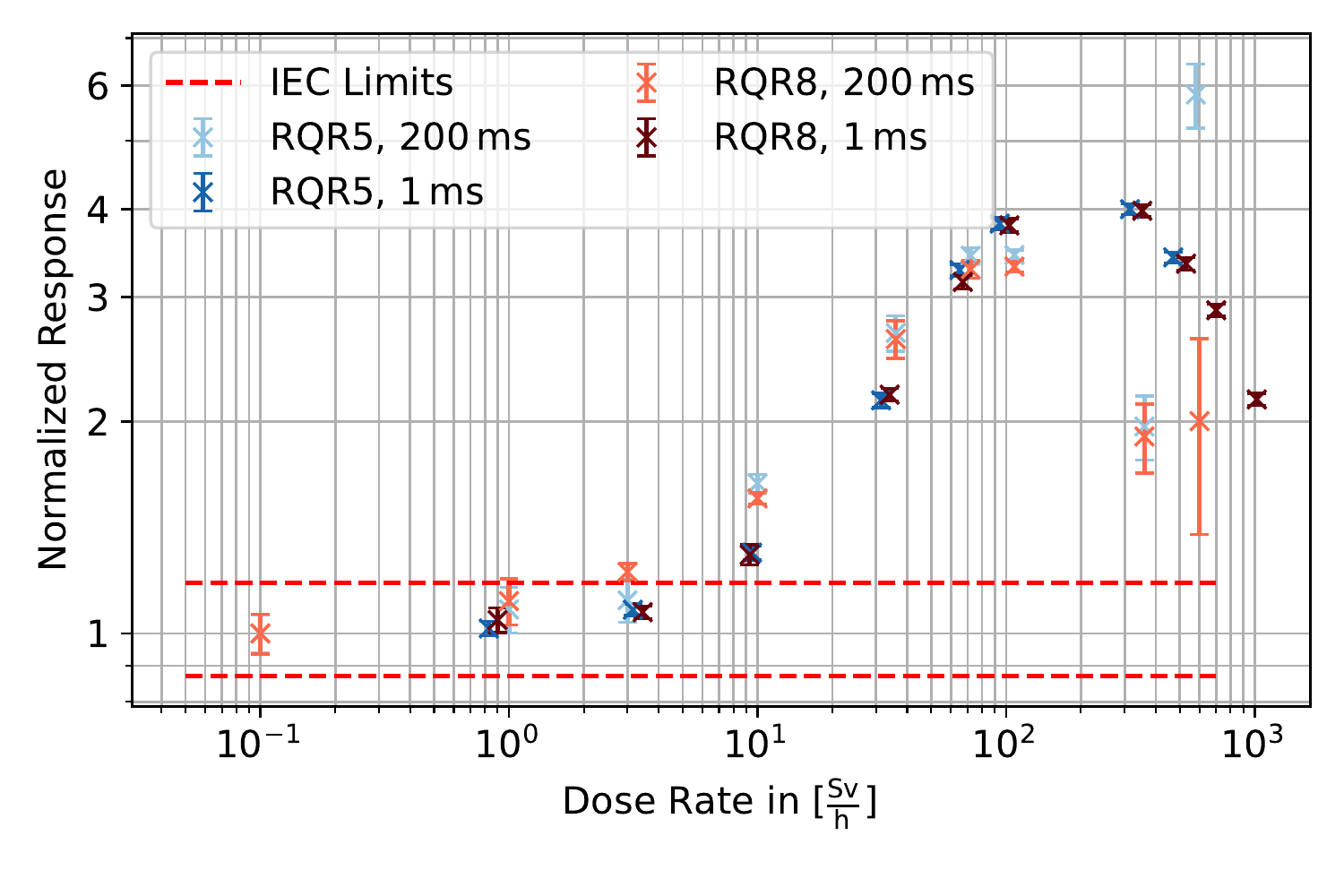}}
    \caption{Normalized response of the eye lens dosemeter prototype with respect to irradiated dose rate \(\Dot{H}_\text{p}(3)\) for radiation qualities RQR-5 and RQR-8 and for pulse durations $200\,$ms and $1\,$ms. Normalization was performed with respect to the measurement of the lowest dose rate for each quality. The uncertainty bars represent the error of the mean value derived from four individual measurements per dose rate. The dashed red lines indicate the approval limits of 0.87 to 1.18, stated in IEC 61526 \cite{iec61526} for dose rates up to \(1\,\frac{\text{Sv}}{\text{h}}\).}
    \label{fig:Rate}
\end{figure}
The resulting normalized response with respect to dose rate is shown in \figurename~\ref{fig:Rate}.
For both radiation qualities, response normalization was performed with respect to the measurement at \(0.1\,\frac{\text{Sv}}{\text{h}}\).
As stated in IEC 61526 \cite{iec61526} and the extension of PTB-A 23.2 \cite{ptba23e}, the normalized response must not fall below 0.87 or exceed 1.18 for dose rates \(\Dot{H}_\text{p}(3)\leq1\,\frac{\text{Sv}}{\text{h}}\).
This is fulfilled for both tested radiation qualities and pulse durations. 
Starting at \(4\,\frac{\text{Sv}}{\text{h}}\), dose is increasingly overestimated.
This is caused by increasing contributions of analog pile-up events as the average time distance of two signals in one pixel starts to decrease for higher dose rates.
Two photons interacting in a sensor pixel with a time distance shorter than the process time of the first signal cannot be distinguished from each other.
Rather, they are treated as a single energy deposition with larger energy resulting in distortions of the measured energy deposition spectra and, consequently, results in a wrongly estimated dose.
On the other hand, neither the radiation quality nor the pulse duration is observed to have a significant influence on the dose estimation behavior of Dosepix with respect to dose rate.

Furthermore, the influence of the pulse duration on the response of the detector was investigated.
Irradiations were performed with an RQR-8 reference photon field at a constant dose rate of \(4\,\frac{\text{Sv}}{\text{h}}\) while varying the radiation pulse duration.
Each measurement was repeated four times.
\begin{figure}
    \centerline{\includegraphics[width=3.5in]{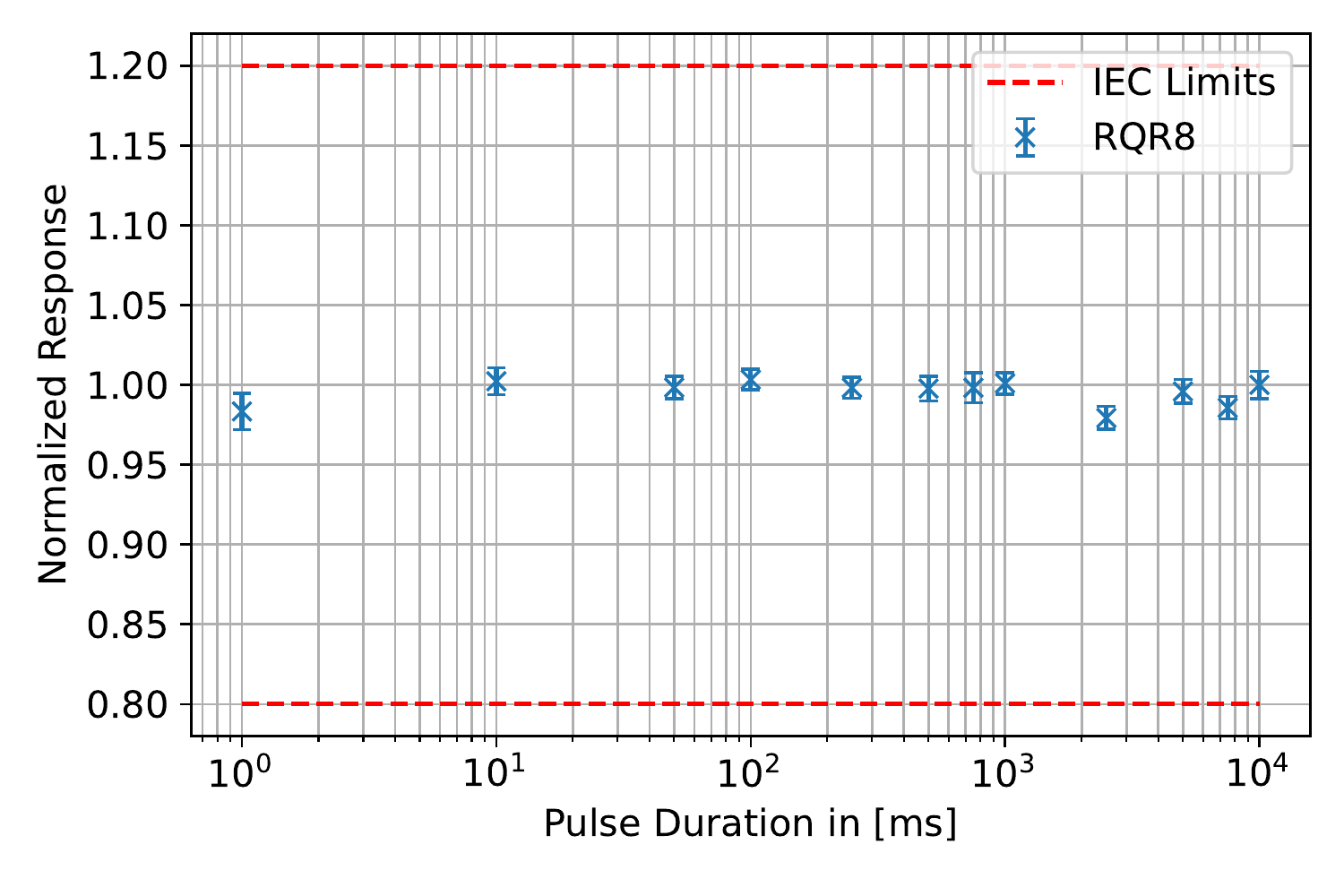}}
    \caption{Normalized response of the eye lens dosemeter prototype with respect to pulse duration $T$. Normalization of the response was performed with respect to the continuous measurement at \(T=10\,\)s. The uncertainty bars represent the error of the mean value derived from four individual measurements per pulse duration. The red dashed lines indicate approval limits stated in IEC 62743 \cite{iec62743} of a maximum deviation of the normalized response of \(\pm 20\,\%\).}
    \label{fig:PAbh}
\end{figure}
Results are shown in \figurename~\ref{fig:PAbh} together with the approval limits, as stated in IEC 62743~\cite{iec62743}, marked as red dashed lines.
The shortest pulse duration of \(1.0\,\)ms is given by technical limits of the x-ray facility whereas the longest value of 10\,s is given by the defined transition from pulsed to continuous radiation \cite{iec62743}.
The results are normalized to the response value at continuous radiation, i.e., 10\,s.
No significant change of the normalized response with respect to the pulse duration can be observed.
An increase of the statistical fluctuations for shorter pulses results from the small number of registered events.

\subsection{Reproducibility}

Reproducibility of the measured dose of the active eye lens dosemeter prototype was tested by applying selected radiation qualities (N-30, N-80, N-120, N-150, and N-300) four times for ten seconds each.
The relevant parameter for reproducibility is the coefficient of variation $v$ which is defined as
\begin{equation}
    v=\frac{\sigma}{\mu}
\end{equation}
where $\sigma$ is the standard deviation of a set of measurements and $\mu$ their mean value.
\begin{figure}
    \centerline{\includegraphics[width=3.5in]{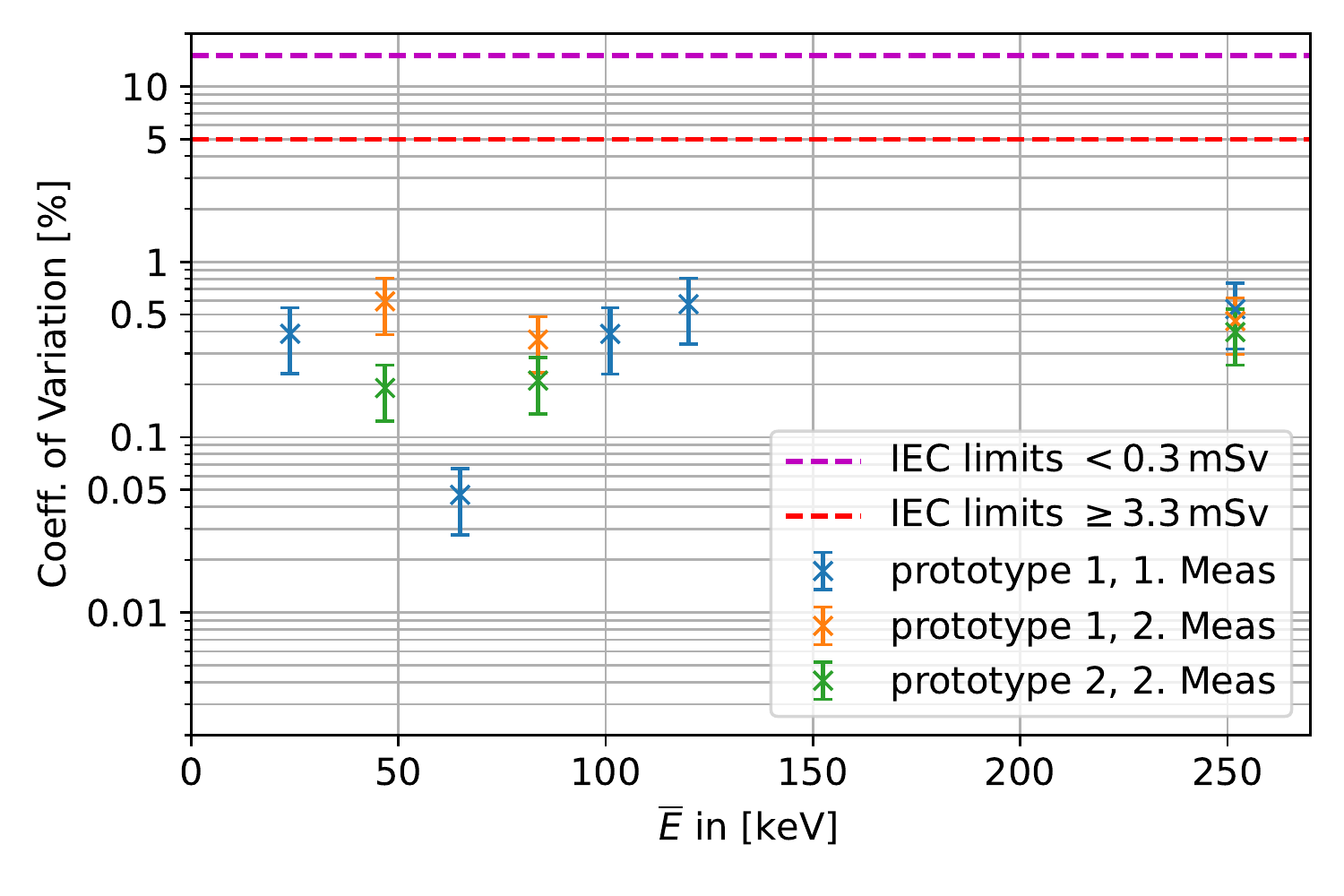}}
    \caption{Coefficient of variation for different radiation qualities, calculated from four identical measurements per quality with a reference dose of \(168\,\upmu\)Sv (1. measurement) or \(3.4\,\)mSv (2. measurement) indicated in the legend. The dashed lines indicate $15\,\%$ and $5\,\%$ which are the approval limits for \(H_\text{p}(3)<0.3\,\)mSv and \(H_\text{p}(3)\geq3.3\,\)mSv stated in IEC 61526 \cite{iec61526}. }
    \label{fig:Rep}
\end{figure}
In \figurename~\ref{fig:Rep}, the measured coefficients of variation for all tested radiation qualities are shown.
Two reference dose values were chosen.
The first series of measurements were performed with a reference dose of \(H_\text{p}(3)_\text{ref}=168\,\upmu\)Sv whereas the second series was performed with \(H_\text{p}(3)_\text{ref}=3.4\,\)mSv.
Only the latter was repeated with dosemeter prototype 2 because the first series was performed before prototype 2 was assembled. 
A red and a magenta dashed line mark the approval maximum variations of \(15\,\%\) and \(5\,\%\) which are allowed for \(H_\text{p}(3)<0.3\,\)mSv and \(H_\text{p}(3)\geq 3.3\,\)mSv respectively according to IEC 61526 \cite{iec61526} and PTB-A 23.2 \cite{ptba23e}.
$168\,\upmu$Sv were chosen for the same reason as $167\,\upmu$Sv above with the difference originating from systematic uncertainties of the setup.
The high reference dose of 3.4\,mSv was chosen as it is slightly above aforementioned 3.3\,mSv stated in \cite{iec61526}.
All estimated coefficients of variation stay below \(1\,\%\), i.e., they are far below the approval limits.
According to these results, the coefficient of variation does not significantly depend on the reference dose which is proportional to the number of expected events in the detector.
Therefore, variation is not dominated by statistical fluctuations due to event statistics.
There is also practically no energy dependence of the measured coefficient of variation - apart from statistical fluctuations – as can be seen for the measurement of N-80 with a mean energy of 65\,keV originating from the small number of measurements.
Details on the fluctuation of a measured coefficient of variation can be found in the literature \cite{Brunzendorf}.
To check these results, the variation of indicated dose should be measured for a series of reference doses to further elaborate on its dose dependence, the reference dose should especially be reduced down to a value where statistical uncertainties start to dominate the results.
A second tested active eye lens dosemeter prototype shows similar behavior while even slightly outperforming the first prototype in terms of reproducibility.
Its results are also shown in \figurename~\ref{fig:Rep} affirming the ability to produce comparable prototypes.
However, this has to be proven by manufacturing and testing further devices. 

\section{Comparison with Other Dosemeters}

A reliable comparison of the results with other dosemeters is difficult as, at the time of writing, there are no legally approved active eye lens dosemeters.
Therefore, a comparison has to be made with active whole-body dosemeters and passive eye lens dosemeters.
The dosiEYE system by Mirion Technologies \cite{Hoedlmoser} and the EYE-D dosemeter by RADCARD \cite{Bilsky} are two available passive eye lens dosimetry systems.
Both are able to measure \(H_\text{p}(3)\) in the energy range from N-20 (\(\overline{E}=16.3\,\)keV) to N-300 (\(\overline{E}=248\,\)keV) as well as S-Cs (\(\overline{E}=662\,\)keV) and S-Co (\(\overline{E}=1250\,\)keV) with a response deviation within the approval limits.
The response of EYE-D also does not deviate more than \(\pm20\,\%\) for angles of incidence up to \(\pm 75^\circ\) \cite{Bilsky}.
Passive dosemeters are also insensitive to different dose rates \cite{Karsch}.
However, the presented active eye lens dosemeter prototype holds up well to the passive systems between 12\,keV and 250\,keV and for dose rates \(\leq1\,\frac{\text{Sv}}{\text{h}}\).

For comparison with active dose measuring systems, the Electronic Personal dosemeter Thermo Scientific EPD Mk2+ by the company Thermo Scientific Fisher \cite{epd} and RaySafe i3 by Fluke Biomedical \cite{raysafe} are considered.
Both are whole-body dosemeters which measure the personal dose equivalent \(H_\text{p}(10)\).
Their energy range of use reaches up to 1250\,keV (S-Co).
However, they are only able to measure the dose within \(\pm 20\,\%\) uncertainty up to \(300\,\frac{\text{mSv}}{\text{h}}\) (RaySafe i3) or \(1\,\frac{\text{Sv}}{\text{h}}\) (EPD Mk2+).
These limits are surpassed or, at least, caught up by the presented active eye lens dosemeter prototype.
Dose estimation of the active eye lens dosemeter varies less (\(<1\,\%\)) than both of the considered whole-body dosemeter (\(\approx 10\,\%\)) i.e. the results of the former are more reproducible.
In summary, the presented measurements show that an active eye lens dosemeter based on the hybrid pixelated detector Dosepix has the potential to be, at least, in line with current passive eye lens dosemeters and active whole-body dosemeters – with the very large benefit of active warning functions in case of high dose (rate) values.

\section{Conclusion and Outlook}

The first prototype of an active personal eye lens dosemeter based on Dosepix is able to excellently measure the personal dose equivalent \(H_\text{p}(3)\).
Dose estimation within the approval limits for such dosemeters has been proven successfully for N-series radiation qualities between N-15 to N-300 and angles of radiation incidence from $-75^\circ$ to $75^\circ$ in continuous photon fields.
The dose from pulsed photon fields is correctly measured for pulse durations down to \(1\,\)ms and dose rates up to \(1\,\frac{\text{Sv}}{\text{h}}\) for radiation qualities RQR-5 and RQR-8.
Its reproducibility plainly fulfills the approval limits stated in IEC 61526 \cite{iec61526} and the extension of PTB-A 23.2 \cite{ptba23e}.

Future investigations must focus on the behavior of the eye lens dosemeter under realistic conditions.
At the time of writing, all of the presented tests have been performed at PTB under fixed laboratory conditions and with a well-defined setup.
However, the real life scenario for this dosemeter mainly lies in monitoring dose exposures of medical staff in interventional radiology and cardiology, thus studies under these conditions are crucial for a final evaluation of the usability.
This includes dose estimation of scattered x-radiation with and without the available protection measures as well as long term studies.

Also, the prototype has to be transformed into its final form.
Currently, data transfer and power consumption are realized via a USB-connection between the dosemeter and a PC.
Dose calculation and data analysis is exclusively executed at a PC which is not convenient for the targeted application.
The final dosemeter is going to be powered by a rechargeable battery.
Dose is going to be calculated by the microcontroller inside the dosemeter.
There will also be a visual display on the dosemeter case and data transfer with a tablet PC via Bluetooth.

\section*{Acknowledgements}
This work was funded by \textit{Bayerisches Staatsministerium für Wirtschaft, Landesentwicklung und Energie} under grant M4-1906-0006 within the scope of \textit{Medical Valley Award}, 2018 (project title: Electronic X-ray Tracker).
All authors declare that they have no known conflicts of interest in terms of competing financial interests or personal relationships that could have an influence or are relevant to the work reported in this paper.
The authors are grateful to Simone Janßen and Katharina Olzem (both PTB) for their valuable support during the measurements.


\end{document}